\documentclass[onecolumn]{revtex4}

\usepackage{graphicx}
\usepackage{dcolumn}
\usepackage{bm}

\input epsf

\begin{document}

\title{\Large TRAJECTORY AROUND A SPHERICALLY SYMMETRIC NON-ROTATING BLACK HOLE}

\author{\bf Sumanta
~Chakraborty** , Subenoy
Chakraborty\footnote{schakraborty@math.jdvu.ac.in }}

\affiliation{ **Department ~Of ~Physics,~ Presidency~
College,~Kolkata,~India,
 *Department ~Of ~Mathematics,~Jadavpur~
University,~Kolkata,~India.}

\date{\today}

\begin{abstract}
Trajectory of a test particle or a photon around a general
spherical black hole is studied and bending of light trajectory is
investigated. Pseudo-Newtonian gravitational potential describing
the gravitational field of the black hole is determined and is
compared with the related effective potential for test particle
motion. As an example, results are presented for
Reissner-Nordstr\"{o}m
 black hole.
\end{abstract}

\pacs{04.50.-h, 04.40.Dg, 97.60.Lf}

\maketitle

\section{Introduction}
Motion of a test particle around a black hole is a very old topic
of investigation to know the behavior of the gravitational field
around black hole(for example [1-3] and any other book on general
relativity). Such investigation was started long back when within
a year of publication of general theory of relativity by
Einstein(1915)[4-7], Schwarzchild [8] gave a vacuum solution to
the Einstein field equations. The solution describes the geometry
of vacuum space-time outside a spherical massive body and is known
today as Schwarzchild black hole solution. At present in all text
books on general relativity there is an exhaustive study of the
motion of a test particle around the Schwarzchild black hole(see
for example [1-3]). In the present work, an investigation of the
motion of a test particle is done around a general static
non-rotating black hole. A general formula for determining bending
of light is evaluated and is tested for Schwarzchild black
hole.Gravitational field outside the black hole is approximated by
Pseudo-Newtonian(PN) gravitational potential and is compared with
the corresponding effective potential for test particle motion.
Finally all results are verified for
Reissner-Nordstr\"{o}m black hole solution.\\

The paper is organized as follows: Section II deals with the
motion of a massive test particle around a general spherically
symmetric non-rotating black hole. Also effective potential,energy
and condition for circular orbit are determined for the black
hole. Further some comments are presented point wise. In section
III trajectory of a photon and bending of light is studied.
Pseudo-Newtonian gravitational potential is determined and it is
compared with effective potential in section IV. As an example all
results are deduced for
Reissner-Nordstr\"{o}m black hole in section V. The paper ends with a short conclusion in section VI. \\

\section{Motion Of a Test Particle around A General Black Hole}
 The line element of a static
spherically symmetric space time which describes a black hole can
be written as,
\begin{equation}
ds^{2}=-f(r)dt^{2}+\frac{dr^{2}}{f(r)}+r^{2}d\Omega_{2}^{2}
\end{equation}
where $f(r)$ is at least a $c^{2}$ function and it should satisfy the following conditions so that line element (eq.1) describes a black hole solution: i) $f(r)$ must have a zero at some positive r (say $r_{h}$) so that time dilation is infinite at $r_{h}$, ii) The Kretschmann scalar ($\alpha=R^{ijkl}R_{ijkl}$) should be finite at $r=r_{h}$ but it diverges at $r=0$ i.e. the space time described by equation (1) has curvature singularity only at $r=0$. Here
$$d\Omega_{2}^{2}=d\theta^{2}+\sin^{2}\theta d\phi^{2}$$is the metric on unit two sphere.
 Suppose we consider the motion of a
test particle of rest mass m around the black hole. The
corresponding Lagrangian will be
\begin{equation}
2L=-f(r)\left(\frac{dt}{d\lambda}\right)^{2}+f(r)^{-1}\left(\frac{dr}{d\lambda}\right)^{2}+
r^{2}\left(\frac{d\theta}{d\lambda}\right)^{2}+r^{2}\sin^{2}\theta\left(\frac{d\phi}{d\lambda}\right)^{2}
\end{equation}
where $\lambda$ is any affine parameter.\\

As the Lagrangian has two cyclic co-ordinates $t$ and $\phi$ so
the corresponding momenta must be constant. This leads to
\begin{equation}
E=-\frac{p_{0}}{m}
\end{equation}
a constant and
 \begin{equation}
 L=\frac{p_{\phi}}{m}
\end{equation}
 is also a constant.

As the space-time is spherically symmetric so the motion is always
confined to a plane which for convenience chosen to be the
equatorial plane $\theta=\frac{\pi}{2}$. The explicit form of the
momentum components are
$$p^{~0}=E\frac{m}{f(r)}$$
$$p^{~r}=m\frac{dr}{d\lambda}$$
\begin{equation}
 p^{~\theta}=0
\end{equation}
$$ p^{~\phi}=\frac{mL}{r^{2}}$$
Using the above expressions for momentum components in the energy
momentum conservation relation
\begin{equation}
p^{\mu}p_{\mu}=-m^{2}
\end{equation}
 we obtain
\begin{equation}
\left(\frac{dr}{d\lambda}\right)^{2}=E^{2}-V^{2}(r)
\end{equation}
Where
\begin{equation}
V^{2}(r)=f(r)\left(1+\frac{L^{2}}{r^{2}}\right)
\end{equation}
is called the (square of the) effective potential.\\

Now differentiating both sides of equation (7) we have
\begin{equation}
\frac{d^{2}r}{d\lambda^{2}}=-\frac{1}{2}\frac{dV^{2}(r)}{dr}
\end{equation}
Also from (5) the momentum in the $\phi$ direction gives
\begin{equation}\frac{d\phi}{d\lambda}=\frac{L}{r^{2}}
\end{equation}
So eliminating the affine parameter between $(7)$ and $(10)$ the
differential equation of the trajectory of the particle in the
equatorial plane is given by
\begin{equation}
\left(\frac{dr}{d\phi}\right)^{2}=\frac{r^{4}}{L^{2}}\left[E^{2}-f(r)\left(1+\frac{L^{2}}{r^{2}}\right)\right]
\end{equation}
Which can be written as
$$\left(\frac{dr}{d\phi}\right)^{2}=\frac{r^{4}}{L^{2}}\psi(r)$$
Where
\begin{equation}
\psi(r)=E^{2}-f(r)\left(1+\frac{L^{2}}{r^{2}}\right)
\end{equation}

We can make the following conclusions on the trajectory of the
particle:\\

$\bullet~~~$ The energy of the particle should not be less than
the potential $V(r)$ i.e. for a given $E$ the trajectory should be
such that the radial range is restricted to those radii for which
$V$ is smaller than $E$.\\

$\bullet~~~$ If $\psi(r)>0$ for all values of $r$ then the
particle comes from infinity and moves directly to the origin.
This is called the terminating escape orbit.\\

$\bullet~~~$ If $\psi(r)$ has one positive zero then the particle
either starts from finite distance moves directly to the origin
(known as terminating bound orbit) or it may move on an escape
orbit with a finite impact parameter $\frac{L}{E}$.\\

$\bullet~~~$ If $\psi(r)$ has two positive zeros then we have two
possible cases: I. if $\psi(r)>0$ between the two zeros then the
trajectory is called periodic bound orbit like planetary orbit or
II. if $\psi(r)<0$ between these two zeros then the trajectory is
either an escape orbit or a terminating bound orbit.\\

$\bullet~~~$ The points where $\psi(r)=0$ are known as turning
points of the trajectory i.e. the value of r which satisfies
\begin{equation}
E^{2}=f(r)\left(1+\frac{L^{2}}{r^{2}}\right)
\end{equation} are turning points and Eq.(13) determines the potential
curves.\\

$\bullet~~~$For circular orbit $ \left(r=constant\right)$ we have
from (9)$\frac{dV^{2}(r)}{dr}=0$ i.e. circular orbits are possible
for those radial co ordinates which correspond to maximum
(unstable) or minimum(stable) of the potential. Thus for circular
orbit we must have
\begin{equation}
\frac{f'(r)}{f(r)}=\frac{2L^{2}}{r(r^{2}+L^{2})}
\end{equation}

\section{Trajectory of a photon: Bending of Light}

To determine the photon trajectory we shall proceed as before.
Here from the energy momentum conservation relation we have

\begin{equation}
\left(\frac{dr}{d\lambda}\right)^{2}=E^{2}-f(r)\frac{L^{2}}{r^{2}}
\end{equation}
i.e. $V_{L}(r)^{2}=f(r)\frac{L^{2}}{r^{2}}$ is the (square of the)
effective potential. So differentiating both sides we
get\begin{equation}
 \frac{d^{2}r}{d\lambda^{2}}=-\frac{1}{2}\frac{dV_{L}^{2}(r)}{dr}
\end{equation}
Thus the differential path of a light ray is given by
\begin{equation}
\frac{d\phi}{dr}=\pm\frac{1}{r^{2}\sqrt{\left[\frac{1}{b^{2}}-\frac{1}{r^{2}}f(r)\right]}}
\end{equation} where $b=\frac{L}{E}$. Now for photon circular orbit we have $$\frac{dv^{2}_{L}}{dr}=0$$ i.e. the radius of the circular orbit satisfies
$$rf'(r)=2f(r)$$ One may note that the radius of photon circular orbit is independent of the angular momentum of photon.
In particular, for a Schwarzchild black hole the radius of photon
circular orbit is $3M$ .For an ingoing photon choosing
$u=\frac{1}{r}$ we have
\begin{equation}
\frac{d\phi}{du}=\frac{1}{\sqrt{\left[\frac{1}{b^{2}}-u^{2}F(u)\right]}}
\end{equation}
 where $F(u)=f(\frac{1}{u})$.\\

Note that if $F(u)$ is a constant then $(18)$ has the solution
\begin{equation}
r\sin(\phi-\phi_{0})=b
\end{equation} (choosing the constant to be unity) a straight line. This is expected as f(r)=constant
means the space time is minkowskian (having no gravitational
effect) and photon trajectory will be straight line. Further we
see that at large distance (small $u$) the gravitational field due
to the black hole will be negligible so we may expand $F(u)$ in a
power series of u i.e.
\begin{equation}
F(u)=1+c_{1}u+c_{2}u^{2}+\cdots
\end{equation}
So keeping up to first order in $u$ we have from
$(18)$\begin{equation}
\frac{d\phi}{du}=\frac{1}{\sqrt{\left[\frac{1}{b^{2}}-u^{2}-c_{1}u^{3}\right]}}
\end{equation} Let $y=u\left(1+\frac{c_{1}u}{2}\right)$ then
$u=y\left(1-\frac{c_{1}y}{2}\right)+0\left(u^{2}\right)$. Then the
above differential equation becomes
\begin{equation}
\frac{d\phi}{dy}=\frac{1-c_{1}y}
{\sqrt{\frac{1}{b^{2}}-y^{2}}}+0\left(u^{2}\right)
\end{equation} So on integration,
\begin{equation}
\phi=\phi_{0}-\frac{c_{1}}{b}+\sin^{-1}{(by)}+c_{1}\sqrt{\frac{1}{b^{2}}-y^{2}}
\end{equation} If we choose $\phi_{0}$ as the initial incoming
direction of light i.e. $\phi\longrightarrow\phi_{0}$ as
$y\longrightarrow0$ and as in the approximation $y=\frac{1}{b}$
corresponds to the smallest $r$ that photon can travel then
\begin{equation}
\phi_{y=\frac{1}{b}}=\phi_{0}-\frac{c_{1}}{b}+\frac{\pi}{2}
\end{equation} So the angle of deflection is
$\delta=\frac{\pi}{2}-\frac{c_{1}}{b}$, as the photon comes from
infinity to the point of closest approach. Hence the total
deflection would be $\delta=\pi-\frac{2c_{1}}{b}$. Therefore,
considering the straight line path the net amount of deflection
will be
\begin{equation}
\Delta \phi=-\frac{2c_{1}}{b}
\end{equation}
 If we take schwarzschild solution then we have
$c_{1}=-2M$ and $\Delta\phi=\frac{4M}{b}$. Further, if it so
happen that $c_{1}=0$ then we choose $F(u)\simeq1+c_{2}u^{2}$.
Using the transformation
$$y=u\left(1+\frac{c_{2}u^{2}}{2}\right)$$ the net amount of
deflection will be given by
\begin{equation}
\Delta \phi=\frac{3 \pi c_{2}}{4b^{2}}
\end{equation}

\section{Pseudo-Newtonian gravitational and effective potentials}
Based on a general heuristic method([9]) ,the PN gravitational
potential can be defined as
\begin{equation}
\psi=\int\frac{l_{c}^{2}}{r^{3}}dr
\end{equation}
where, r is the usual radial co-ordinate and $l_{c}$ is the
general relativistic specific angular momentum i.e.
$l_{c}(=\frac{L_{c}}{E_{c}})$ is the ratio of the conserved
angular momentum and energy per particle mass, related to the
circular geodesic in the equatorial plane.\\

 In newtonian theory, the gravitational potential $\psi_{n}$ is
 given by $$\psi_{n}=\int\frac{l_{cn}^{2}}{r^{3}}dr$$ with
 $l_{cn}$, the newtonian angular momentum per mass of the particle moving in the circular orbit.
 The motivation of choosing PN potential (27) is to match the
 newtonian angular momentum per particle mass on a circular orbit
 with the general relativistic angular momentum.
        In the present study, the general relativistic conserved
angular momentum and energy per particle mass for circular orbit
are given by (from equations (13) and (14))
\begin{equation}
L_{c}=\left[\frac{r^{3}f'(r)}{2f(r)-rf'(r)}\right]^{\frac{1}{2}}
\end{equation}
and
\begin{equation}
E_{c}=\frac{\sqrt{2}f(r)}{\sqrt{\left[2f(r)-rf'(r)\right]}}
\end{equation}
i.e.
\begin{equation}
l_{c}=\frac{1}{f(r)}\sqrt{\frac{r^{3}f'(r)}{2}}
\end{equation}
Hence from (27) the PN gravitational potential is
\begin{equation}
\psi=c-\frac{1}{2f(r)}
\end{equation}
where the constant of integration 'c' is determined from the known
result of the Schwarzchild black hole as follows:(note that c has
no physical
meaning)\\

 For Schwarzchild black hole, the well known
 Paczy\'{n}ski-Witta gravitational potential([10]) is
\begin{equation}
\psi_{PW}=-\frac{M}{r-2M}
\end{equation}
substituting in eq.(31) we get $c=\frac{1}{2}$ and hence the PN
gravitational potential for a general spherically symmetric black
hole described by eq.(1) is
\begin{equation}
\psi=\frac{1}{2}\left[1-\frac{1}{f(r)}\right]
\end{equation}
As for static radius $(r_{s})$ gravitational potential should be
zero so from (33) we have $f(r_{s})=1$. Hence from equations
(28),(29) the circular orbit of the test particle exist for radius
$(r_{c})$ in the range([11])
$$r_{a}<r_{c}<r_{s}$$
where $r_{a}$ satisfies $$rf'(r)='2f(r)$$ i.e. $r_{a}$ is the
photon circular orbit. Thus all circular orbits of the test
particle are lower bounded by the photon circular orbit and are
extended upto the static radius.\\

 Further, from eq.(33) we see that the PN potential diverges at
 the event horizon (i.e.$f(r)=0$) reaches its maximum value at
 $r=r_{m}$ (where$f'(r_{m})=0$) and then decreases for
 $r>r_{m}$ i.e. the gravitational field corresponding to PN
 potential become repulsive for $r>r_{m}$. Also if the metric(1)
 becomes asymptotically flat(i.e. $f(r)\longrightarrow1$ as
 $r\longrightarrow\infty$) then $\psi\longrightarrow0$
 asymptotically([11],[12]).
 Moreover, for central gravitational fields, if we assume that the
 motion of the test particle is confined to the equatorial plane
 then for Keplarian motion along the radial direction gives
\begin{equation}
\frac{1}{2}\left(\frac{dr}{dt}\right)^{2}=e-v_{eff}
\end{equation}
where e stands for total PN energy per particle mass and $v_{eff}$
stands for PN effective potential per particle mass having
explicit form ([13])
\begin{equation}
v_{eff}=\psi+\frac{l^{2}}{2r^{2}}
\end{equation}
where $\psi$ is the PN gravitational potential given by eq.(33)
and $l$ is the PN angular momentum per particle mass.
 Thus circular Keplarian orbits are characterized by the extrema
 of
 $v_{eff}$(i.e.$\frac{dv_{eff}}{dr}=0$) and we have
\begin{equation}
l_{c}^{2}=\frac{r^{3}f'(r)}{2\left[f(r)\right]^{2}}
\end{equation}
which can be written as (using eq.s (28) and (29))
\begin{equation}
l_{c}=\frac{L_{c}}{E_{c}}
\end{equation}
The corresponding expression for energy is
\begin{equation}
e_{c}=\frac{1}{2}\left[1+\frac{rf'(r)-2f(r)}{2\left(f(r)\right)^{2}}\right]
\end{equation}
using $E_{c}$ it can be written as
\begin{equation}
e_{c}=\frac{1}{2}\left[1-\frac{1}{E_{c}^{2}}\right]
\end{equation}
It is to be noted that angular momentum per particle mass is same
for both General Relativity as well as for PN effective potential
theory.\\

 We shall now examine the stability of circular orbits studied
 above. As stability criteria is determined by the extrema of the
 effective potential $v_{eff}$, so we have
 $$\frac{\partial l_{c}^{2}}{\partial r}>0$$ for stable circular
 orbit and $$\frac{\partial l_{c}^{2}}{\partial r}<0$$ for unstable
 circular orbit. Due to identical nature of $l_{c}$ both GR and PN
 potential theory have same criteria for stability. Further, inner
 and outer marginally stable circular orbits corresponds to
 extrema of $l_{c}^{2}$ and we have
\begin{equation}
2r\left[f'(r)\right]^{2}=f(r)\left[3f'(r)+rf''(r)\right]
\end{equation}
Hence for stability one should have
\begin{equation}
2r \left[f'(r) \right]^{2}<f(r) \left[3f'(r)+rf''(r) \right]
\end{equation}
Finally, from the above expressions (eq.s (36) and (38)) of
$l_{c}^{2}$ and $e_{c}$ we have the following observations:\\

$\bullet~~~~~$ At the event horizon($f(r)=0$) both $l_{c}^{2}$ and
$e_{c}$ diverge while $E_{c}$ and $L_{c}$ exist if
$[2f(r)-rf'(r)]$ is positive definite. For example in case of
Schwarzchild-de Sitter space time([11]) they become
finite while for our Reissner-Nordstr\"{o}m black hole (see next section) they do not exist.\\

$\bullet~~~~~$ At the static radius ($f(r)=1$)  both $l_{c}^{2}$
and $e_{c}$ vanish while $E_{c}$ and $L_{c}$ depend on choice of
$f(r)$.\\

$\bullet~~~~$ At the photon circular orbit ($rf'(r)=2f(r)$) both
$L_{c}$ and $E_{c}$ diverge while $e_{c}=\frac{1}{2}$ and $l_{c}$
is finite there.\\

   Thus we conclude that both in PN approach and in relativistic
approach the circular orbits are bounded from above by static
radius while bound from below by event horizon in PN approach and
that by the radius of photon orbit in relativistic approach. This
is due to the fact that we do not get photon circular orbit in PN
approach([11]).\\

\section{An Example: Reissner-Nordstr\"{o}m Black Hole}
For Reissner-Nordstr\"{o}m(R-N) black hole solution we have
\begin{equation}
f(r)=1-\frac{2M}{r}+\frac{q^{2}}{r^{2}}
\end{equation}
where M being the mass and q the charge of black hole. So for
horizon we have $f(r)=0$ i.e.
\begin{equation}
r_{\pm}=M\pm\sqrt{M^{2}-q^{2}}
\end{equation}
Black hole solution exists for $$M^{2}>q^{2}$$ and we have event
horizon at $$r_{h}=r_{+}$$ and $$r_{c}=r_{-}$$ corresponds to
black hole Cauchy horizon. When $$M^{2}=q^{2}$$ then both the
horizons coincide and it is the case of extremal black hole.

The static radius is given by ($f(r)=1$)
\begin{equation}
r_{s}=\frac{q^{2}}{2M}
\end{equation}
 The radius at which $f'(r)=0$gives the solution$$r_{m}=\frac{q^{2}}{M}$$
One may note that $$r_{s}<r_{m}<r_{+}$$ So both the static radius
and $r_{m}$ lie inside the event horizon. Hence they have no
physical significance for R-N black hole.

Also the radius of photon circular orbits are given by
\begin{equation}
r_{pc}=\frac{1}{2}\left[3M\pm\sqrt{9M^{2}-8q^{2}}\right]
\end{equation}

\textbf{Relativistic Theory}:\\

For circular orbit the conserved angular momentum and energy per
particle mass are
\begin{equation}
L_{c}=\frac{r\sqrt{Mr-q^{2}}}{\sqrt{r^{2}-3Mr+2q^{2}}}
\end{equation}
Variation of $L_{c}$ with respect to the variation of $q$ and $r$
for $M=1$ has been shown in figure 1 [variables in the figures are in any standard units].

\begin{figure}
\includegraphics[height=3in, width=3in]{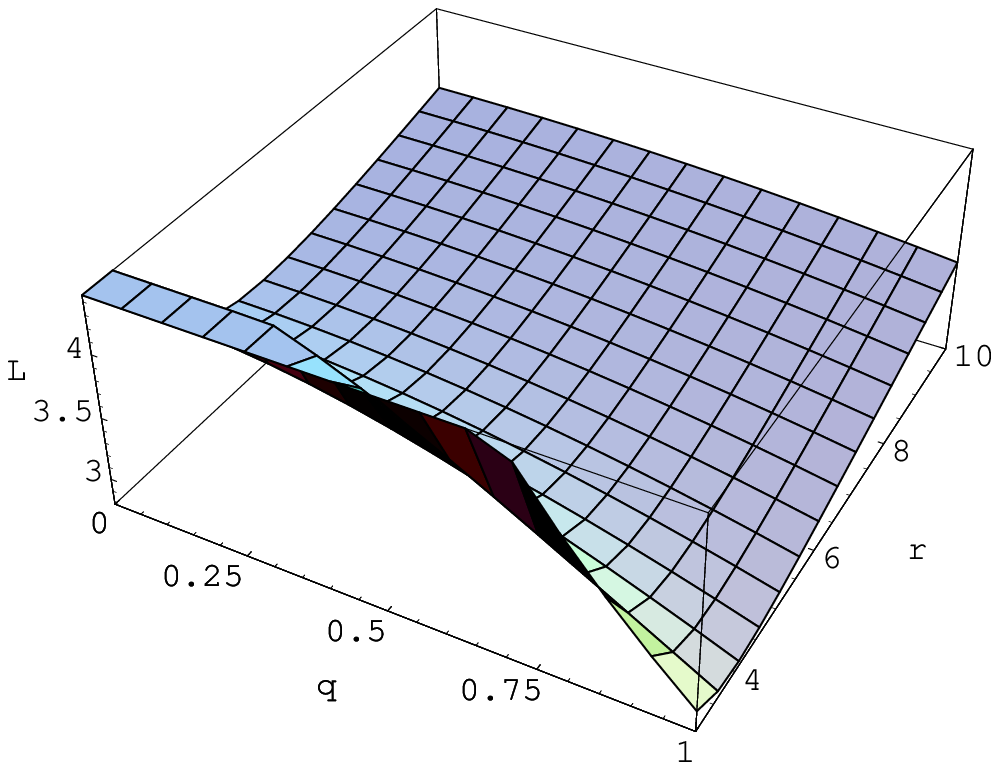}~~~

\vspace{1mm}

Figure 1

\vspace{6mm}Figure 1: The figure shows the variation of
relativistic conserved angular momentum $L_{c}$ with respect to
variation of q and r for the choice $M=1$.\hspace{1cm}
 \vspace{6mm}
\end{figure}

\begin{equation}
E_{c}=\frac{r\left[1-\frac{2M}{r}+\frac{q^{2}}{r^{2}}\right]}{\sqrt{r^{2}-3Mr+2q^{2}}}
\end{equation}
\begin{equation}
l_{c}=\frac{L_{c}}{E_{c}}=\frac{\sqrt{Mr-q^{2}}}{1-\frac{2M}{r}+\frac{q^{2}}{r^{2}}}
\end{equation}
with effective potential
\begin{equation}
V_{eff}=\sqrt{\left[\left(1-\frac{2M}{r}+\frac{q^{2}}{r^{2}}\right)\left(1+\frac{L^{2}}{r^{2}}\right)\right]}
\end{equation}

\textbf{Pseudo-Newtonian Theory}:\\

PN gravitational potential and effective potentials are
\begin{equation}
\psi=\frac{q^{2}-2Mr}{2\left[r^{2}-2Mr+q^{2}\right]}
\end{equation}
The graph of $\psi$ for variation of both q and r with $M=1$ is
presented in figure 2. Also in figure 3, we have shown the
dependence of $\psi$ by drawing the graphs of $\psi$ for three
different values of q.

\begin{figure}
\includegraphics[height=3in, width=3in]{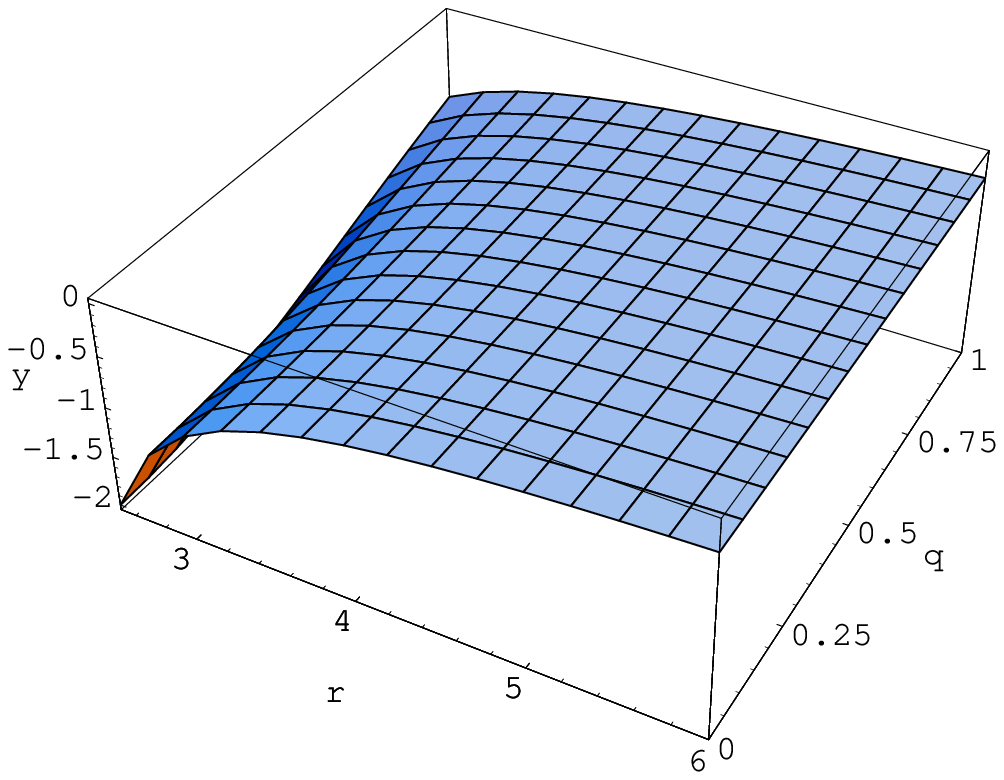}~~~

\vspace{6mm}

Figure 2

\vspace{6mm}Figure 2: Here variation of PN gravitational potential
$\psi$ with variation of
 both q and r for $M=1$ is shown.\hspace{1cm}
\vspace{6mm}
\end{figure}

\begin{figure}
 \includegraphics[height=3in, width=3in]{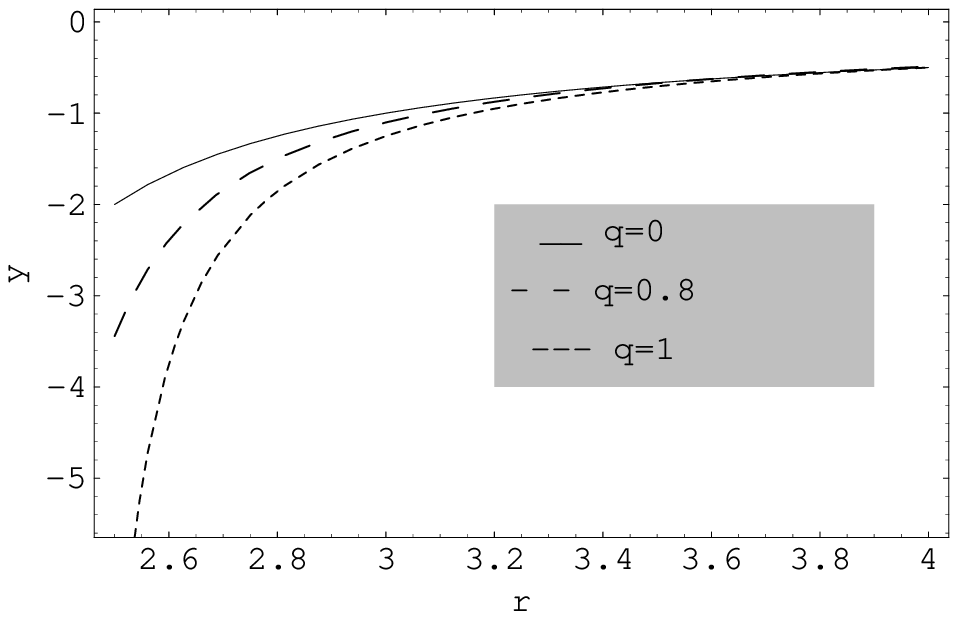}~~~

\vspace{6mm}

Figure 3

 \vspace{6mm}Figure 3:Here PN gravitational potential $\psi$ is drawn for three different values of q. The upper one for $q=0$, middle one for $q=0.8$
  and the lower one for $q=1$. \hspace{1cm}
 \vspace{6mm}
\end{figure}

\begin{equation}
v_{eff}=\frac{q^{2}-2Mr}{2\left[r^{2}-2Mr+q^{2}\right]}+\frac{l^{2}}{2r^{2}}
\end{equation}
Also we have
\begin{equation}
l_{c}=\frac{\sqrt{Mr-q^{2}}}{1-\frac{2M}{r}+\frac{q^{2}}{r^{2}}}
\end{equation}
The graphical presentation of $l_{c}$ for the variation of both q
and r with $M=1$ has been shown in figure 4. Also graphically a
comparative study of $L_{c}$(given by eq.(46)) and $l_{c}$(given
by eq.(52)) has been presented in figure 5(a) and 5(b) for two
different values of q.

\begin{figure}
 \includegraphics[height=3in, width=3in]{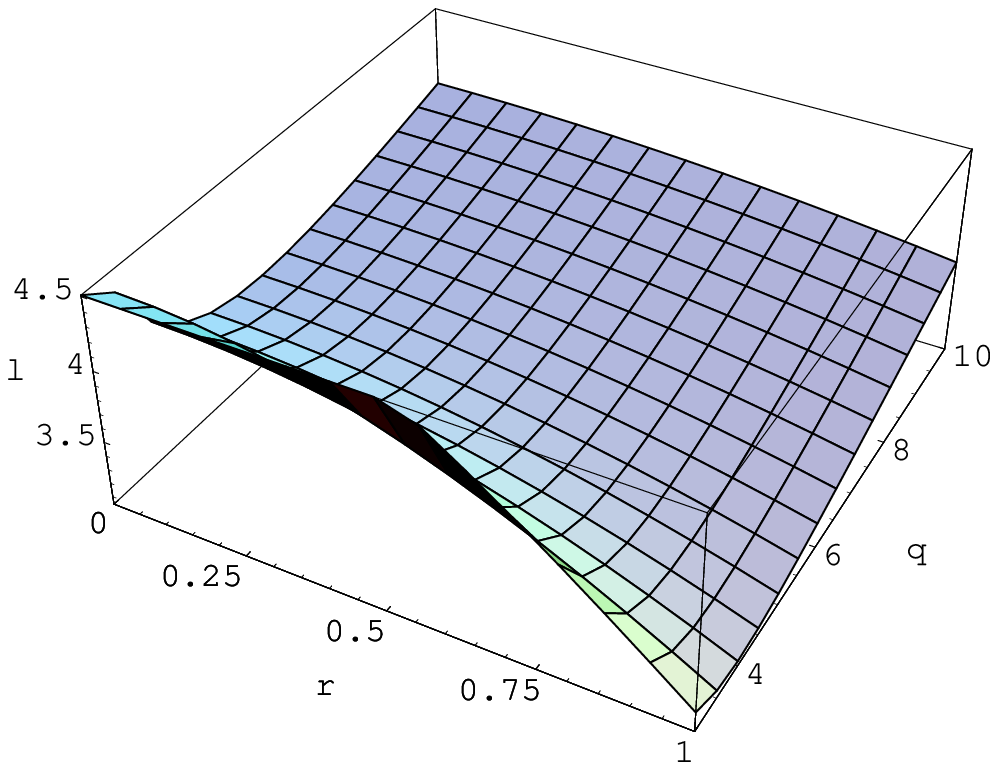}~~~

\vspace{6mm}

Figure 4

 \vspace{6mm}Figure 4:The figure shows the graphical representation of PN angular momentum per particle mass
 $l_{c}$ for variation of both q and r with $M=1$ \hspace{1cm}
 \vspace{6mm}
\end{figure}

\begin{figure}
\includegraphics[height=3in, width=3in]{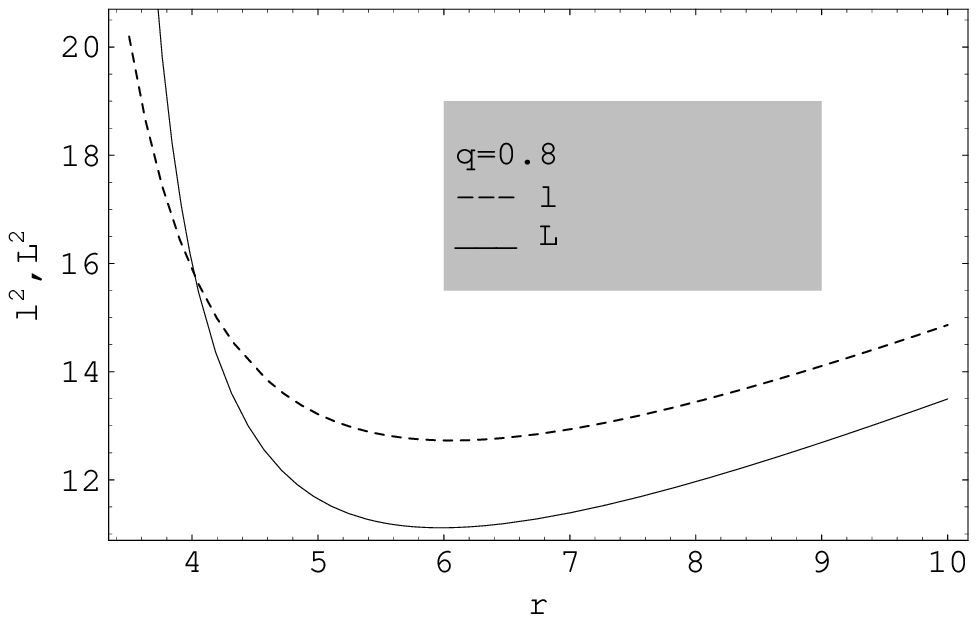}~~~
\includegraphics[height=3in, width=3in]{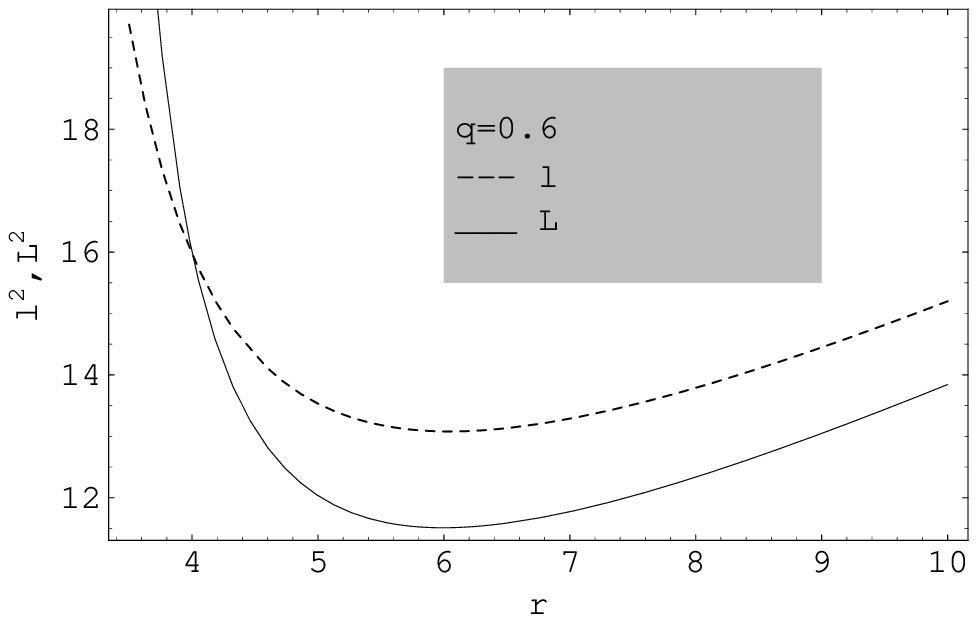}~~~

\vspace{6mm}

Figure 5(a) ~~~~~~~~~~~~~~~~~~~~~~~~~~~~Figure 5(b)

 \vspace{6mm}Figure 5(a) and 5(b) show the comparative study of the variation of relativistic angular momentum
 $L_{c}$(given by eq.(46)) and PN angular momentum per particle mass $l_{c}$(given by eq.(52))
 for $q=0.8$ and $0.6$ respectively.Where upper one gives $L_{c}$ and lower one $l_{c}$  \hspace{1cm}
 \vspace{6mm}
\end{figure}

\begin{equation}
e_{c}=\frac{q^{4}-Mr^{3}-4Mrq^{2}+4M^{2}r^{2}}{2\left[r^{2}-2Mr+q^{2}\right]^{2}}
\end{equation}
The marginally stable circular orbits are given by the positive
root of the equation
\begin{equation}
Mr^{3}-6M^{2}r^{2}+9Mq^{2}r-4q^{4}=0
\end{equation}
If the cubic eq has three positive real
roots($r_{m1}<r_{m2}<r_{m3}$)then $r=r_{m3}$ and $r=r_{m1}$ are
respectively the radii of the outer and inner marginally stable
circular orbits.

Finally, for stable circular orbit we have
\begin{equation}
Mr^{3}-6M^{2}r^{2}+9Mq^{2}r-4q^{4}>0
\end{equation}

\section{Conclusion}

In this work we give a general formulation of the trajectory of a
test particle (or a photon) around any spherically symmetric black hole in four
dimensional space time. We also classify the trajectories by
studying the possible positive zeros of the function
$$\psi(r)=E^{2}-f(r)\left(1+\frac{L^{2}}{r^{2}}\right)$$ So once
$f(r)$ is given for a given black hole we can immediately tell the
trajectories of a test particle or a photon around it. In the PN
approach physical parameters for circular orbits are evaluated and
are compared with the corresponding quantities in relativistic
treatments. Stability condition for the circular orbits are
determined and bounds of the marginally stable circular orbits are
compared in both formalism. As an example we have applied our
results for R-N black hole solution.\\

It is to be noted that the above analysis of the trajectory is not
restricted to Einstein gravity, it can also be applied to any
black hole solution in a modified gravity theory. Further our
analysis can be extended to any higher dimension. For example, the
metric ansatz for a $n$ dimensional black hole is written in the
form
\begin{equation}
ds^{2}=-f(r)dt^{2}+\frac{dr^{2}}{f(r)}+r^{2}d\Omega_{n-2}^{2}
\end{equation}
 Where $d\Omega_{n-2}^{2}$ is the metric on unit
$n-2$ sphere and is given by $$d\Omega_{1}^{2}=d\phi^{2},$$
$$d\Omega_{i+1}^{2}=d\theta_{i}^{2}+\sin^{2}\theta_{i}d\Omega_{i}^{2}, i\geq1 $$
 Then due to spherical symmetry of the space time the motion of a
test particle can be restricted to the equatorial plane defined by
$\theta_{i}=\frac{\pi}{2}, i\geq1$. As before energy $E$ and the
angular momentum $L$ are two conserved quantities and the
differential equation for the path of the test particle becomes
\begin{equation}
\left(\frac{dr}{d\phi}\right)^{2}=\frac{r^{4}}{L^{2}}\left[E^{2}-f(r)\left(\epsilon+\frac{L^{2}}{r^{2}}\right)\right]
\end{equation}
where $\epsilon=0$ or $1$ for photon or massive particle. Finally, one may note that throughout our calculations we have not used any specific gravity theory. So if we have a black hole solution given by equation (1) not only in Einstein gravity but also in any other gravity theory, the above analysis of the trajectory of a test particle is valid. Moreover from equations (56) and (57), we may conclude that above analysis of particle trajectory can be extended to any dimension of space time.\\
          For future work, an extension of the above approach to non-spherical systems (particularly axi-symmetric) would be interesting.\\

{\bf Acknowledgement:}\\

The first author is thankful to Dr.Prabir Kr. Mukherjee,
Department of Physics, Presidency College, Kolkata, for valuable
help in preparing the manuscript.The first author is also thankful
to DST, Govt.of India for awarding KVPY fellowship.The authors
greatly acknowledge the warm hospitality at IUCAA, Pune where
apart of the
work has been done.\\\\

{\bf References:}\\

 $[1]$. Narlikar.J.V {\it Lectures on General Relativity and Cosmology}
 (The Macmillan Company of India) (1978)\\

$[2]$ Schutz.Bernerd {\it A First Course in General Relativity}
(Cambridge University Press)(1995)\\

$[3]$ Ray.D.Inverno {\it Introducing Einstein's  General Theory of
Relativity} (Clarenden Press,Oxford)(2003) \\

$[4]$ Einstein.A {\it Preuss, Akad.Wiss.Berlin, Sitzber,} \textbf{778},\textbf{799},\textbf{831} and \textbf{844} (1915)\\

$[5]$ Einstein.A {\it Preuss, Akad.Wiss.Berlin, Sitzber,} \textbf{142} (1917) \\

$[6]$ Einstein.A {\it The Meaning of Relativity}(Methuen, London) (1951) \\

$[7]$ Einstein. A {\it Relativity: the special and general theory}
(Methuen, London) (1920)\\

$[8]$ Schwarzchild.K  {\it Sitzber, Deut.Akad.Wiss.Berlin.KI.Math-Phys. Tech.} \textbf{189} (1923) \\

$[9]$ Mukhopadhyay.B {\it Astrophysical Journal} {\bf
581} (2002) 427.\\

$[10]$ Paczy\'{n}ski.B and Witta.p {\it Astron.Astrophys.}
\textbf{88} (1980) 23.\\

$[11]$ Stuchl\'{i}k.Z and Kov\'{a}\v{r}.J {\it Int.J.Mod.Phys.D}
\textbf{17} (2008) 2089.\\

$[12]$ Stuchl\'{i}k.Z, Slan\'{y} and Kov\'{a}\v{r}.J {\it
Class.Quantum.Grav} \textbf{26} (2009) 215013.\\

$[13]$ Minser.C.W, Thorne.K.S and Wheeler.J.A, {\it Gravitation}
(Freeman, San Francisco, 1973).\\

\end{document}